\documentclass[english,pra,twocolumn]{revtex4-1}
\usepackage[T1]{fontenc}
\usepackage[latin9]{inputenc}
\usepackage{color}
\usepackage{mathrsfs}
\usepackage{amsmath}
\usepackage{amssymb}
\usepackage{graphicx}

\makeatletter

\providecommand{\tabularnewline}{\\}

\usepackage[colorlinks = true,
            linkcolor =red,
            urlcolor  =magenta,
            citecolor = blue,
            anchorcolor = blue]{hyperref}

\makeatother

\usepackage{babel}
\begin{document}
\title{Stronger sum uncertainty relations for non-Hermitian operators }
\author{Xiao-Feng Song,Yi-Fang Ren, Shuang Liu, Xi-Hao Chen}
\author{Yusuf Turek}
\email{yusufu1984@hotmail.com}

\affiliation{$^{1}$School of Physics, Liaoning University, Shenyang, Liaoning
110036, China}
\date{\today}
\begin{abstract}
The uncertainty relations (URs) of two arbitrary Hermitian and non-Hermitian
incompatible operators represented by the product of variances have
been confirmed theoretically and experimentally in various physical
systems. However, the lower bound of the product uncertainty inequality
can be null even for two non-commuting operators, i.e., a trivial
case. Therefore, for two incompatible operators over the measured
system state, the associated URs regarding the sum of variances are
valid in a state-dependent manner, and the lower bound is guaranteed
to be nontrivial. Although the sum URs formulated for Hermitian and
unitary operators have been affirmed, the general forms for arbitrary
non-Hermitian operators have not yet been investigated. This study
presents the sum URs for non-Hermitian operators acting on system
states using an appropriate Hilbert-space metric. The compatible forms
of our sum inequalities with the conventional quantum mechanics are
also provided via the $G$-metric formalism. Concrete examples illustrate
the validity of the proposed sum URs in both $\mathcal{PT}$-symmetric
and $\mathcal{PT}$-broken phases. The developed methods and results
can help give an in-depth understanding of the usefulness of $G$-metric
formalism in non-Hermitian quantum mechanics and the sum URs of incompatible
operators within.
\end{abstract}
\maketitle

\section{Introduction}

The uncertainty principle (UP) and uncertainty relations (URs) are
crucial in quantum mechanics, offering insights into the behavior
of microscopic systems. Initially formulated by Heisenberg \citep{1927Heisenberg},
UP provides a fundamental constraint on the simultaneous measurement
precision of canonically conjugate observables, setting a lower bound
on their product of errors and disturbances. The Heisenberg uncertainty
principle is a cornerstone in quantum mechanics, fundamentally affecting
the understanding of physical systems by establishing intrinsic limits
on the precision of certain pairs of observables. However, Heisenberg's
original statement referred to the error and disturbance in a measurement
process, which depends on the measurement techniques. With the development
of quantum measurement theory and the related techniques, recently,
the incorrectness \citep{RevModPhys.42.358}, modifications \citep{PhysRevA.67.042105,OZAWA2004350,Mir_2007,Lund_2010},
and the violations \citep{2004v,PhysRevLett.109.100404,2013B} of
Heisenberg\textquoteright s measurement-disturbance relationship due
to weak measurements were investigated. It should be noted that Heisenberg's
original proposal was different compared to its current interpretation.
Kennard \citep{1927KennardZurQE} and Weyl \citep{1928H} provided
the textbook forms of the position and momentum of UR based on their
variances. Roberston \citep{PhysRev.34.163} rigorously established
the Heisenberg UR for pairs over the state $\vert\psi\rangle$ as

\begin{equation}
\Delta A^{2}\Delta B^{2}\geq\frac{1}{4}\left|\langle\psi\vert\left[A,B\right]\vert\psi\rangle\right|^{2},
\end{equation}
where $\triangle A^{2}$ and $\triangle B^{2}$ are the variances
of non-commuting Hermitian operators, defined as $\triangle A=\sqrt{\langle A^{2}\rangle-\langle A\rangle^{2}}$
and $\triangle B=\sqrt{\langle B^{2}\rangle-\langle B\rangle^{2}}$,
and $\langle X\rangle=\langle\psi\vert X\vert\psi\rangle$ is the
average of an operator $X$ in state $\vert\psi\rangle$. $\left[A,B\right]=AB-BA$
denotes the commutator of $A$ and $B$. Unlike Heisenberg's original
measurement-disturbance relationship, the Roberston UR presented above
is independent of any specific measurement. Notably, the inequality
may become trivial even when considering that $A$ and $B$ are incompatible
with some system states, which can be derived by applying the Cauchy-Schwarz
inequality. In 1930, Schrödinger \citet{1930Schrodinger} improved
this relation by adding an expectation value term of an anti-commutator
$\left\{ A,B\right\} $ on the right-hand side (RHS). However, this
improved form still suffered from the trivial cases mentioned above. 

In addition to using variance to characterize the URs adopted by Roberston
and Schrödinger, another widely recognized approach to describing
the URs is through entropy, which was first proposed by Deutsch \citep{PhysRevLett.50.631}
and later optimized by Maassen and Uffink \citep{PhysRevLett.60.1103}.
Additionally, many researchers developed various uncertainty relations
based on different entropy measures \citep{PhysRevA.74.052101,PhysRevA.79.062108,PhysRevLett.106.110506,PhysRevA.102.022217}.
In addition to the entropy method, there are a host of methods interpreting
URs via different forms, e.g., in terms of noise and disturbance \citep{PhysRevLett.111.160405},
successive measurement \citep{PhysRevLett.50.631,PhysRevA.87.062112},
majorization technique \citep{2013Puchala,PhysRevLett.111.230401},
and skew information \citep{PhysRevLett.91.180403,2021Zhang,2022Ma}.
This product of variances URs was tested experimentally in various
aspects \citep{2011Li,2012Erhart,PhysRevA.88.022110,2021Dengke} within
conventional quantum mechanics (CQM) formalism. URs are beneficial
for a wide range of applications, including quantum teleportation
\citep{PhysRevA.86.032338}, quantum steering \citep{PhysRevA.87.062103},
quantum key distribution \citep{PhysRevA.89.022112,PhysRevA.93.042324},
quantum foundation \citep{2010Jonathan,PhysRevLett.129.240401}, quantum
random number generation \citep{PhysRevA.90.052327,PhysRevX.6.011020},
entanglement detection \citep{PhysRevA.90.062127,PhysRevLett.112.180501,2020Zhang},
quantum spin squeezing \citep{PhysRevLett.47.709,1985Wodkiewicz,PhysRevA.46.R6797,PhysRevA.47.5138,2011Jian},
quantum metrology \citep{PhysRevLett.72.3439,1996Samuel,PhysRevLett.96.010401,2011Giovannetti,2015Jarzyna},
quantum cryptography \citep{PhysRevA.53.2038}, quantum gravity \citep{2004Michael},
and quantum information science \citep{2006Masato,2010Berta,2016Majumdar,2019Yunger}. 

Most URs are formulated in CQM, where the operators are assumed to
be Hermitian. Nevertheless, Hermiticity is an axiom of quantum mechanics
that guarantees probability preservation and a real spectrum. In 1998,
Bender \citep{PhysRevLett.80.5243} proved that the strict Hermiticity
requirement for a system to have a real spectrum can be replaced with
the less restrictive condition of $\mathcal{PT}$ symmetry. In a $\mathcal{PT}$-symmetric
system, its eigenspectrum is real even though its corresponding Hamiltonian
is non-Hermitian, which has gained a broad interest for non-Hermitian
quantum mechanics (NHQM). In $\mathcal{PT}$-invariant non-Hermitian
systems \citep{PhysRevLett.89.270401,PhysRevA.100.062121,2020T.Ohlsson},
a transition occurs that divides them into two phases: one where the
system exhibits $\mathcal{PT}$ symmetry with a completely real spectrum,
and another where $\mathcal{PT}$ symmetry is broken, leading to a
spectrum comprising complex conjugate pairs, either entirely or partially.
In NHQM, the product of variances UR for two non-Hermitian operators
$A$ and $B$ is expressed as \citep{PhysRevA.92.052120} 
\begin{equation}
\triangle A^{2}\triangle B^{2}\ge\vert\langle A^{\dagger}B\rangle-\langle A^{\dagger}\rangle\langle B\rangle\vert^{2},\label{eq:2-3}
\end{equation}
where the variance of the operator $\mathcal{O}=A,B$ is defined as
$\triangle\mathcal{O}^{2}=\langle\mathcal{O^{\dagger}O}\rangle-\langle\mathcal{O^{\dagger}}\rangle\langle\mathcal{O}\rangle$.
In recent years, URs were also investigated in NHQM theoretically
\citep{PhysRevA.92.052120,PhysRevA.93.052118,2022Zhao,2023Bagarello,PhysRevA.107.042201}
and experimentally \citep{PhysRevLett.120.230402,PhysRevLett.132.070203}. 

In CQM and NHQM, most URs are based on the product of variances $\Delta A^{2}\Delta B^{2}$
of the observables. However, those products of variances can be zero
even if one of the variances is nonzero, which is trivial. Therefore,
these earlier uncertainty relations fail to completely capture the
incompatibility among the observables in the system state. Interestingly,
recent studies focused on sum URs since they were nontrivial whenever
the operators are incompatible with the state. In Ref. \citep{PhysRevLett.113.260401},
Maccone and Pati established a pair of URs for sums of variances in
CQM, consistently yielding nontrivial bounds even in the case of observable
eigenstates. The stronger URs were tested experimentally by the outcomes
of the projective measurements to obtain every term directly \citep{PhysRevA.93.052108}.
In Ref. \citep{PhysRevA.94.042104}, the sum URs for general unitary
operators were investigated in detail, which was experimentally demonstrated
for two three-level unitary operators with photonic qutrits \citep{2017Lei}.
In open quantum systems various non-Hermitian operators exist, such
as unitary operators, ladder operators, and effective non-Hermitian
Hamiltonians. Thus, the sum URs investigated in Ref. \citep{PhysRevA.94.042104}
only considered the special case of sum URs in NHQM. Additionally,
that study did not consider the $G$-metric formalism of NHQM. Furthermore,
multiple observables URs were proposed in related theoretical works
\citep{2015Chen,2016qiu,2016Chen} and experimental works \citep{PhysRevA.96.062123,PhysRevA.98.032118,2019Chen}. 

This study investigates the sum URs in non-Hermitian systems by considering
the $G$-metric formalism. As investigated in Refs. \citep{PhysRevLett.112.130404,PhysRevA.90.054301,2014Pati},
directly applying the axioms and theorems of CQM to non-Hermitian
systems may conflict with certain theoretical principles, which are
fundamental in quantum physics. Therefore, a modified non-Hermitian
quantum mechanics (NHQM) formulation was developed \citep{PhysRevA.100.062118,Tu2023generalpropertiesof,Ju2024emergentparallel}
based on Hilbert space geometry. This formulation is consistent with
the CQM for Hermitian systems by employing the Hermitian positive
definite matrix $G$ alongside generalized operators to ensure the
probability is time-invariant. In the $G$-metric formalism of NHQM,
for every nonzero $\vert\psi(t)\rangle$, one always chooses $\langle\psi(t)\vert G(t)\vert\psi(t)\rangle>0$,
so that $G$ is positive definite and the probability is time invariant.
For a given Hamiltonian, different metrics $G(t)$ exist that are
related by a covariantly constant transition function. Different choices
of a metric $G$ correspond to different choices of bases, which are
not limited to the eigenkets of the Hamiltonian. Thus, if the corresponding
$G$ uses the eigenstates, which form a complete set of bases, the
$G$ is the same metric in biorthogonal quantum mechanics \citep{2013Brody}.
It was shown that the NHQM does not violate the theorems in CQM, including
the no-go theorems, if the state and adjoint operators are modified
by $G$-metric construction. Additionally, unlike the Dirac inner
product used in CQM, $\mathcal{PT}$-symmetric quantum theory can
effectively utilize the $G$-inner product. Indeed, CQM emerges as
a particular instance within the $\mathcal{PT}$-symmetric NHQM, specifically
under the $G$-metric inner product formalism. As a fundamental cornerstone
of quantum mechanics, URs should also preserve their validity in NHQM.
In Ref. \citep{PhysRevA.107.042201}, the researchers presented a
very elegant form of product variances UR of two incompatible non-Hermitian
operators using the $G$-metric formalism. Nevertheless, its corresponding
sum URs have not yet been investigated. Therefore, reassessing the
sum URs of NHQM by considering the $G$-metric formalism is mandatory.

This paper investigates the nontrivial lower bound for the sum of
the variances applied to general non-Hermitian operators in NHQM.
Additionally, this work provides rigorous proof for it. Then, the
bounds of the proposed sum URs are strengthened within the $G$-inner
product framework. We also employ the general good observable condition
to construct the modified compatible forms of URs with CQM. Furthermore,
we observe similarities in their mathematical forms by comparing the
derived URs with their counterparts in CQM. The theoretical results
are verified numerically by employing two distinct examples. 

The rest of this paper is organized as follows. Section \ref{sec:2}
provides explicit details on the derivations of the four sum URs developed
for arbitrary two non-Hermitian operators and presents the modified
forms in terms of $G$-metric formalism. Section \ref{sec:3}, presents
two different examples to prove the validity of our sum URs in all
NHQM realms, including $\mathcal{PT}$-symmetric and $\mathcal{PT}$-symmetry
broken phases, and discusses the results. Section \ref{sec:4}, gives
some discussions and concludes this work.

\section{\protect\label{sec:2} Sum URs for non-hermitian operators}

This section introduces four URs for two incompatible non-Hermitian
operators. These relations are discussed within the NHQM framework
and in its $G$-metric formalism. This section also provides the form
of these inequalities under the condition of good observables \citep{PhysRevA.107.042201}.

\textit{First and second inequalities:} Let $A$ and $B$ be two non-Hermitian
operators in a Hilbert space $\mathcal{H}$, i.e., $A^{\dagger}\neq A$
and $B^{\dagger}\neq B$. For simplicity, we assume that the operators
in this study are bounded operators \citep{2018Teta,2023Bagarello}.
The scalar product between state vectors $\varphi$ and $\psi$ and
the norm of $\varphi$ can be defined as $\text{\ensuremath{\langle\varphi\vert\psi\rangle=\int\varphi^{\ast}\psi d\tau}}$
and $\vert\vert\varphi\vert\vert=\sqrt{\langle\varphi\vert\varphi\rangle}$,
respectively, with $\varphi,\psi\in$$\mathcal{H}$. Here, $d\tau$
represents an infinitesimal volume associated with our concern. The
self-adjoint (Hermitian conjugate) of an operator $X$ in the Hilbert
space is denoted as $X^{\dagger}$ and defined by $\langle\varphi\vert X\psi\rangle=\langle X^{\dagger}\varphi\vert\psi\rangle$.
Then, for $\hat{X}=X-\langle X\rangle$, one has 
\begin{align}
\vert\vert\hat{X}\varphi\vert\vert^{2} & =\vert\vert(X-\langle X\rangle)\varphi\vert\vert^{2}\nonumber \\
 & =\langle X^{\dagger}X\rangle-\langle X^{\dagger}\rangle\langle X\rangle=\triangle X^{2},\label{eq:2-1}
\end{align}
 where $\langle X\rangle=\langle\varphi\vert X\varphi\rangle=\langle\varphi\vert X\vert\varphi\rangle$
and $\triangle X$ is the standard deviation of the operator $X$
over $\vert\varphi\rangle$.

To present our sum URs for $A$ and $B$ two non-Hermitian operators,
the new operator is assumed to be $X_{\alpha}=\hat{A}-i\alpha\hat{B}$
with $\hat{A}=A-\langle A\rangle$ and $\hat{B}=B-\langle B\rangle,\alpha\in\mathbb{R}$
. From Eq. (\ref{eq:2-1}), one can obtain \citep{2023Bagarello} 

\begin{align}
\vert\vert X_{\alpha}\varphi\vert\vert^{2} & =\alpha^{2}\vert\vert\hat{B}\varphi\vert\vert^{2}+\alpha C_{A,B;\varphi}+\vert\vert\hat{A}\varphi\vert\vert^{2},\label{eq:1}
\end{align}
where $\vert\vert\hat{A}\varphi\vert\vert^{2}=\triangle A^{2}$ and
$\vert\vert\hat{B}\varphi\vert\vert^{2}=\triangle B^{2}$, and \textbf{$C_{A,B;\psi}=-i\langle\hat{A}^{\dagger}\hat{B}-\hat{B}^{\dagger}\hat{A}\rangle$}
is real, satisfied for all values of $\alpha$. Since $\vert\vert X_{\alpha}\varphi\vert\vert^{2}\ge0$,
if $\alpha=-1$, the sum of the variances of $A$ and $B$ obeys the
UR presented below 

\begin{align}
\Delta A^{2}+\Delta B^{2} & \geq2Im\left[Cov\left(A,B\right)\right].\label{eq:2}
\end{align}
Here, the covariance of $A$ and $B$ is defined as $Cov(A,B)=\langle\hat{A}^{\dagger}\hat{B}\rangle=\langle A^{\dagger}B\rangle-\langle A^{\dagger}\rangle\langle B\rangle$,
and $Im$ denotes the imaginary part of a complex number. Another
new operator can be defined as $Y_{\alpha}=\hat{A}+\alpha\hat{B}$,
which is a linear combination of $\hat{A}$ and $\hat{B}$. For this
operator, the relation presented below is held as
\begin{equation}
\vert\vert Y_{\alpha}\varphi\vert\vert^{2}=\alpha^{2}\vert\vert\hat{B}\varphi\vert\vert^{2}+\alpha D_{A,B;\varphi}+\vert\vert\hat{A}\varphi\vert\vert^{2}\geq0,\label{eq:3}
\end{equation}
where \textbf{$D_{A,B;\psi}=\langle\hat{A}^{\dagger}\hat{B}+\hat{B}^{\dagger}\hat{A}\rangle$},
which has a real value. Notably, Eq. (\ref{eq:3}) is invariant under
any arbitrary value $\alpha$. Specifically, for $\alpha=-1$, the
sum UR becomes

\begin{align}
\Delta A^{2}+\Delta B^{2} & \geq2Re\left[Cov\left(A,B\right)\right].\label{eq:4}
\end{align}
Here, $Re$ denotes the real part of a complex number. Since the RHS
of Eqs. (\ref{eq:2}) and (\ref{eq:4}) are the real and imaginary
parts of the same quantity, we rewrite them as 
\begin{equation}
\Delta A^{2}+\Delta B^{2}\geq2\text{\ensuremath{\max}}\{Re\left[Cov\left(A,B\right)\right],Im\left[Cov\left(A,B\right)\right]\}.\label{eq:7-1}
\end{equation}
This is the proof of our first and second inequalities. 

For a given non-Hermitian system described by the Hamiltonian $H\neq H^{\dagger}$,
the above relations can be expressed in the $G$-metric formalism
\citep{PhysRevA.100.062118}. The $G$ metric is necessary for NHQM
to guarantee the probability conservation in time. Notably, $G(t)$
has to be Hermitian, positive-definite, and satisfy the motion equation
for the conserved probability \citep{PhysRevA.100.062118}:

\begin{equation}
\partial G_{t}(t)=i\left[G(t)H(t)-H^{\dagger}(t)G(t)\right].\label{eq:6}
\end{equation}
Thus, the corresponding $G$ metric always exists for a system's given
Hamiltonian $H(t)$. In this new formalism of NHQM, the state and
adjoint operators have some modifications. For the $G$ metric, the
ket vector $\vert\psi(t)\rangle\rangle$ has no distinction between
the conventional $\vert\psi(t)\rangle$, i.e., $\vert\psi(t)\rangle\rangle=\vert\psi(t)\rangle$.
However, the dual corresponding vectors are not just the Hermitian
conjugate of the conventional vectors but are also subject to a linear
map as $\langle\langle\psi(t)\vert=\langle\psi(t)\vert G(t)$. Hence,
in the $G$-metric formalism, the inner product and expectation value
of $A$ under the state $\vert\psi(t)\rangle$ in CQM, $\langle\psi\vert\psi\rangle$
and $\langle A\rangle=\langle\psi\vert A\vert\psi\rangle$, can be
expressed as
\begin{equation}
\langle\langle\psi\vert\psi\rangle\rangle=\langle\psi\vert G\vert\psi\rangle,
\end{equation}
and 
\begin{align}
\langle A\rangle_{G} & =\langle\langle\psi\vert A\vert\psi\rangle\rangle=\langle\psi\vert GA\vert\psi\rangle,
\end{align}
respectively. From this prospective, metric $G(t)$ is not uniquely
determined for a given system Hamiltonian $H(t)$. Different choices
of $G(t)$ correspond to different choices of bases, which are physically
equivalent. This non-uniqueness of $G(t)$ for a given Hamiltonian
$H(t)$ was also explicitly discussed in Refs. \citep{2009T,2009Kleefeld}.
Specifically, if $\{\vert n(t)\rangle\}$ is any complete set of bases
for the states of an arbitrary Hermitian operator in the Hilbert space
for CQM, its completeness relation can be written as $\sum_{n}\vert n(t)\rangle\langle n(t)\vert=1.$
However, when considering the $G$-metric formalism in NHQM, the above
completeness relation becomes as \citep{Tu2023generalpropertiesof}
\begin{equation}
\sum_{n}\vert n(t)\rangle\rangle\langle\langle n(t)\vert=\sum_{n}\vert n(t)\rangle\langle n(t)\vert G(t)=1.\label{eq:2-2}
\end{equation}
Thus, different metrics $G(t)$ can be found that correspond to different
choices of bases. 

However, in the $G$-metric formalism of NHQM, the adjoint operator
$A^{\dagger}$ of $A$ changed to $A^{\dagger}\rightarrow G^{-1}A^{\dagger}G$
. Thus, in the $G$ metric, the arbitrary new operator $\hat{X}=X-\langle X\rangle$
and its adjoint one are modified to

\begin{align}
\hat{X} & \rightarrow\hat{X}_{G}=X-\langle X\rangle_{G},\label{eq:7}\\
\hat{X}^{\dagger} & \rightarrow\hat{X}_{G}^{\dagger}=G^{-1}X^{\dagger}G-\langle\psi\vert X^{\dagger}G\vert\psi\rangle.\label{eq:8}
\end{align}
Furthermore, in the $G$-metric formalism, the conventional covariance
function $Cov(A,B)$ is modified to $Cov_{g}\left(A,B\right)=\langle\hat{A}^{\dagger}\hat{B}\rangle_{g}=\langle A^{\dagger}GB\rangle-\langle A^{\dagger}G\rangle\langle GB\rangle$.
For this reason, by using the above modifications of the state vector
and adjoint operators corresponding to the $G$-metric formalism in
NHQM, the sum of the variances of $\triangle A^{2}$ and $\triangle B^{2}$
{[}see Eqs. (\ref{eq:2}) and (\ref{eq:4}){]} are modified as 

\begin{align}
\Delta A_{g}^{2}+\Delta B_{g}^{2} & \geq2Im\left[Cov_{g}\left(A,B\right)\right],\label{eq:G1a}
\end{align}
 and 
\begin{equation}
\Delta A_{g}^{2}+\Delta B_{g}^{2}\geq2Re\left[Cov_{g}\left(A,B\right)\right],\label{eq:12-1}
\end{equation}
respectively. Here, the left-hand sides (LHS) terms of the above relations
represent the variances of the incompatible non-Hermitian operators
$A$ and $B$ in the $G$-metric formalism, which are defined as 

\begin{align}
\Delta A_{g}^{2} & =\langle A^{\dagger}GA\rangle-\langle A^{\dagger}G\rangle\langle GA\rangle,\label{eq:15}
\end{align}
 and 
\begin{equation}
\triangle B_{g}^{2}=\langle B^{\dagger}GB\rangle-\langle B^{\dagger}G\rangle\langle GB\rangle,\label{eq:16-2}
\end{equation}
respectively.

In the $G$-metric formalism of quantum mechanics, the Hermiticity
condition on the operator $X$ can be replaced by a more general and
convenient condition, called the `` \textit{good observable}'' \citep{PhysRevA.100.062118}.
Thus, $X$ is a \textit{good observable} if $X^{\dagger}G=GX$. As
investigated in previous studies, the \textit{good observable} can
be either Hermitian or non-Hermitian \citep{PhysRevA.100.062121,PhysRevA.107.042201}.
Using $X^{\sharp}=G^{-1}X^{\dagger}G$, where $\sharp$ stands for
the corresponding adjoint operator of $X$ in NHQM, the generalized
\textquotedblleft Hermitian operators'' in NHQM can be recovered
using $X^{\sharp}=X\Rightarrow X^{\dagger}G=GX$. For all Hermitian
systems, the metric operators can be set to unity. Obviously, all
the Hermitian operators are good observables under the Dirac product
when $G$ is a unit operator, i.e., $G=\mathcal{\mathbb{I}}$, and
in the $\mathcal{PT}$-symmetric phase, the Hamiltonian $H$ of a
given system is also a good observable \citep{PhysRevA.107.042201}.
However, for non-Hermitian systems, determining a good observable
depends on the $G$ metric, which exhibits distinct characteristics
in the $\mathcal{PT}$-symmetric and $\mathcal{PT}$-broken phases.
Thus, if the operators $A$ and $B$ adhere to the condition of good
observables, $A^{\dagger}G=GA$ and $B^{\dagger}G=GB$, the URs in
Eqs. (\ref{eq:G1a}) and (\ref{eq:12-1}) can be further expressed
as 

\begin{equation}
\Delta A_{G}^{2}+\Delta B_{G}^{2}\geq i\langle\left[B,A\right]\rangle_{G},\label{eq:G1b}
\end{equation}
 and 
\begin{equation}
\Delta A_{G}^{2}+\Delta B_{G}^{2}\geq\langle\left\{ A,B\right\} \rangle_{G}-2\langle B\rangle_{G}\langle A\rangle_{G},\label{eq:14}
\end{equation}
respectively. Here, the anti-commutation relation between $A$ and
$B$ is written as $\left\{ A,B\right\} =AB+BA$. Notably, we intentionally
use the uppercase and lowercase letters of $G$ to distinguish whether
the URs under the good observable condition are being applied. Therefore,
the variances of $A$ and $B$ in the $G$ metric {[}see Eqs.(\ref{eq:15})
and (\ref{eq:16-2}){]} after the \textit{good observable }constraints
are given as\textit{ }
\begin{equation}
\Delta A_{G}^{2}=\langle A^{2}\rangle_{G}-\langle A\rangle_{G}^{2},
\end{equation}
 and 
\begin{equation}
\triangle B_{G}^{2}=\langle B^{2}\rangle_{G}-\langle B\rangle_{G}^{2}.
\end{equation}
 The above relations can be used both in the $\mathcal{PT}$-symmetric
and $\mathcal{PT}$-broken phases. In Ref. \citep{PhysRevA.107.042201},
the authors presented the modified Robertson UR using the good observable
condition, with its form being similar to the conventional. This also
proves that the above two modified sum URs have similar forms as CQM.
Further details are provided in the discussion section of this paper.

\textit{Third and fourth inequalities}: Next, we present another two
sum URs for two incompatible operators in NHQM. In quantum mechanics,
the formula below is valid for any kind (Hermitian and non-Hermitian)
of operator $X$ \citep{PhysRevA.41.11}

\begin{align}
X\vert\psi\rangle= & \langle X\rangle\vert\psi\rangle+\triangle X\vert\psi_{X}^{\perp}\rangle.\label{eq:A-V}
\end{align}
This study assumes that $X$ is a non-Hermitian operator. $\triangle X$
is the standard deviation of non-Hermitian operator $X$ defined in
Eq. (\ref{eq:2-1}), and $\langle X\rangle=\langle\psi\vert X\vert\psi\rangle$
is a expectation value of $X$ over $\vert\psi\rangle$. $\vert\psi_{X}^{\perp}\rangle$
represents a state vector, which is orthogonal to $\vert\psi\rangle$
and it depends on the operator $X$. This expression is called the
Aharonov--Vaidman identity. For two incompatible non-Hermitian operators,
$A$ and $B$, then the above expression also applies to their combinations,
$A\pm iB$, thus

\begin{align}
\left(A\pm iB\right)\vert\psi\rangle & =\left(\left\langle A\right\rangle \pm i\left\langle B\right\rangle \right)\vert\psi\rangle+\Delta\left(A\pm iB\right)\vert\psi_{A+iB}^{\perp}\rangle,\label{eq:17}
\end{align}
where $\vert\psi_{A\pm iB}^{\perp}\rangle$ represents the state vector
orthogonal to $\vert\psi\rangle$, which depends on the operators
$A\pm iB$, and $\Delta\left(A\pm iB\right)$ denotes the standard
deviations of the operators $A\pm iB$ over $\vert\psi\rangle$. By
taking the inner product with any vector $\vert\psi^{\perp}\rangle$
orthogonal to $\vert\psi\rangle$, the Eq. (\ref{eq:17}) becomes
as 

\begin{align}
\langle\psi^{\perp}\vert\left(A\pm iB\right)\vert\psi\rangle= & \Delta\left(A\pm iB\right)\langle\psi^{\perp}\vert\psi_{A+iB}^{\perp}\rangle.\label{eq:16}
\end{align}
In the equation above, the contribution of the first term vanishes
due to $\langle\psi^{\perp}\vert\psi\rangle=0$. The squared modulus
of Eq. (\ref{eq:16}) is obtained as

\begin{align}
\left|\langle\psi^{\perp}\vert\left(A\pm iB\right)\vert\psi\rangle\right|^{2} & =\left(\Delta\left(A\pm iB\right)\right)^{2}\left|\langle\psi^{\perp}\vert\psi_{A+iB}^{\perp}\rangle\right|^{2},\label{eq:19}
\end{align}
where $\langle\psi^{\perp}\vert\psi_{A+iB}^{\perp}\rangle$ is the
inner product of two state vectors $\vert\psi^{\perp}\rangle$ and
$\vert\psi_{A+B}^{\perp}\rangle$, and its squared modulus must be
less than 1, i.e., $\left|\langle\psi^{\perp}\vert\psi_{A+iB}^{\perp}\rangle\right|^{2}\leq1$.
Thus,

\begin{equation}
\left[\Delta\left(A\pm iB\right)\right]^{2}\geq\left|\langle\psi^{\perp}\vert\left(A\pm iB\right)\vert\psi\rangle\right|^{2}.\label{eq:18}
\end{equation}
It is easy to see that the $\left[\Delta\left(A\pm iB\right)\right]^{2}$
can be expanded by 

\begin{equation}
\left[\Delta\left(A\pm iB\right)\right]^{2}=\triangle A^{2}+\triangle B^{2}\pm i\langle\hat{A}^{\dagger}\hat{B}\rangle\mp i\langle\hat{B}^{\dagger}\hat{A}\rangle.\label{eq:21}
\end{equation}
By substituting Eq. (\ref{eq:21}) into Eq. (\ref{eq:18}), we obtain
a sum UR as 

\begin{align}
\triangle A^{2}+\triangle B^{2} & \geq\pm2Im\left[Cov\left(A,B\right)\right]+\left|\langle\psi^{\perp}\vert\left(A\pm iB\right)\vert\psi\rangle\right|^{2}.\label{eq:c}
\end{align}
Here, $\left|\langle\psi^{\perp}\vert\left(A\pm iB\right)\vert\psi\rangle\right|^{2}=\left|\langle\psi\vert\left(A^{\dagger}\mp iB^{\dagger}\right)\vert\psi^{\perp}\rangle\right|^{2}$.
This is the third sum UR, and it is valid for all non-Hermitian operators.
The derivation of the relation above assumes that $\vert\psi^{\perp}\rangle$
and $\vert\psi_{A\pm iB}^{\perp}\rangle$ should be simultaneously
orthogonal to $\vert\psi\rangle$. Those states can be chosen using
the Aharonov--Vaidman identity $\vert\psi_{A}^{\perp}\rangle=\hat{A}/\triangle A\text{\ensuremath{\vert\psi\rangle}}$
or $\vert\psi_{B}^{\perp}\rangle=\hat{B}/\triangle B\text{\ensuremath{\vert\psi\rangle}}$
and $\vert\psi_{A\pm iB}^{\perp}\rangle=\left(\hat{A}\pm i\hat{B}\right)/\Delta\left(A\pm iB\right)\vert\psi\rangle$.

Next, the fourth sum UR is provided. Assuming that the sum operator
$A+B$ applies to the Aharonov--Vaidman identity, we have

\begin{align}
\left(A+B\right)\vert\psi\rangle & =\left(\left\langle A\right\rangle +\left\langle B\right\rangle \right)\vert\psi\rangle+\Delta\left(A+B\right)\vert\psi_{A+B}^{\perp}\rangle.\label{eq:A+B}
\end{align}
When the orthogonal state $\langle\psi_{A+B}^{\perp}\vert$ to $\vert\psi\rangle$
is multiplied from the left side of the relation presented above,
the result is obtained as 

\begin{align}
\langle\psi_{A+B}^{\perp}\vert\left(A+B\right)\vert\psi\rangle= & \Delta\left(A+B\right),\label{eq:16-1}
\end{align}
and its squared modulus is read as 

\begin{align}
\left|\langle\psi_{A+B}^{\perp}\vert\left(A+B\right)\vert\psi\rangle\right|^{2} & =\Delta\left(A+B\right)^{2}.\label{eq:25}
\end{align}
Applying the parallelogram law of vectors, $2\Delta A^{2}+2\Delta B^{2}\geq\left[\Delta\left(A+B\right)\right]^{2}$,
which holds for arbitrary operators $A$ and $B$, another sum UR
of the two non-Hermitian operators $A$ and $B$ is obtained as follows

\begin{align}
\Delta A^{2}+\Delta B^{2} & \geq\frac{1}{2}\left|\langle\psi_{A+B}^{\perp}\vert\left(A+B\right)\vert\psi\rangle\right|^{2}.\label{eq:d-1}
\end{align}
The lower bound will not be zero unless it is a special case where
state $\vert\psi\rangle$ is an eigenstate of $A+B$. Based on the
equality $2\Delta A^{2}+2\Delta B^{2}=\left[\Delta\left(A+B\right)\right]^{2}+\left[\Delta\left(A-B\right)\right]^{2}$,
and since $\left[\Delta\left(A\pm B\right)\right]^{2}$ are non-negative,
another inequality can be derived for the minus sign. Therefore, the
fourth inequality is obtained in the following form as

\begin{align}
\Delta A^{2}+\Delta B^{2}\geq & \max\{\frac{1}{2}\left|\langle\psi_{A\pm B}^{\perp}\vert\left(A\pm B\right)\vert\psi\rangle\right|^{2}\}.\label{eq:d}
\end{align}

This is the fourth sum UR, which is also valid for any kind of non-Hermitian
operators. It is observed that there is a finite degree of uncertainty,
except in the trivial scenario where $\vert\psi\rangle$ is an eigenstate
of $A\pm B$, indicating that the RHS effectively measures the incompatibility
between $A$ and $B$ on a given state. The Hermitian counterpart
of the third and fourth URs was investigated in Ref. \citep{PhysRevLett.113.260401},
and the derivation presented in this paper is a generalization of
that relation in the NHQM realm.

Next, the above inequalities of Eqs. (\ref{eq:c}) and (\ref{eq:d})
can be obtained regarding the $G$--metric formalism. By considering
the modifications of the state vector and adjoint operators in the
$G$ metric, the third and fourth sum URs are reformulated into:

\begin{align}
\triangle A_{g}^{2}+\triangle B_{g}^{2}\geq & \pm2Im\left[Cov_{G}\left(A,B\right)\right]+\left|\langle\psi^{\perp}\vert G\left(A\pm iB\right)\vert\psi\rangle\right|^{2},\label{eq:28}
\end{align}

\begin{align}
\triangle A_{g}^{2}+\triangle B_{g}^{2}\geq & \max\left[\frac{1}{2}\left|\langle\psi_{A\pm B}^{\perp}\vert G\left(A\pm B\right)\vert\psi\rangle\right|^{2}\right],\label{eq:29}
\end{align}
where $\triangle A_{g}^{2}$ and $\triangle B_{g}^{2}$ are defined
according to Eqs. (\ref{eq:15}) and (\ref{eq:16-2}), respectively,
and $\left|\langle\psi^{\perp}\vert G\left(A\pm iB\right)\vert\psi\rangle\right|^{2}=\left|\langle\psi\vert\left(A^{\dagger}G\mp iB^{\dagger}G\right)\vert\psi^{\perp}\rangle\right|^{2}$.
Note that $\left|\langle\psi_{A\pm B}^{\perp}\vert G\left(A\pm B\right)\vert\psi\rangle\right|^{2}=\langle\psi_{A+B}^{\perp}\vert G\left(A\pm B\right)\vert\psi\rangle\langle\psi\vert\left(A^{\dagger}\pm B^{\dagger}\right)G\vert\psi_{A+B}^{\perp}\rangle$.
Furthermore, in the context of \textit{good observable} conditions,
the above two inequalities are transformed into:

\begin{align}
\triangle A_{G}^{2}+\triangle B_{G}^{2}\geq & \pm i\langle\left[A,B\right]\rangle_{G}+\left|\langle\psi\vert G\left(A\pm iB\right)\vert\psi^{\perp}\rangle\right|^{2},\label{eq:30}
\end{align}

\begin{align}
\triangle A_{G}^{2}+\triangle B_{G}^{2}\geq & \max\left[\frac{1}{2}\left|\langle\psi_{A\pm B}^{\perp}\vert G\left(A\pm B\right)\vert\psi\rangle\right|^{2}\right].\label{eq:31}
\end{align}
Here, $\left|\langle\psi_{A\pm B}^{\perp}\vert G\left(A\pm B\right)\vert\psi\rangle\right|^{2}=\langle\psi_{A\pm B}^{\perp}\vert G\left(A\pm B\right)\vert\psi\rangle\langle\psi\vert G\left(A\pm B\right)\vert\psi_{A\pm B}^{\perp}\rangle$.
The above inequalities represent the sum URs that are proposed in
the context of the modified NHQM. Notably, the relations presented
in the $G$ metric are more elegant than the general forms and fit
with the inner product and probability conservation in time of NHQM. 

\section{\protect\label{sec:3}Examples}

This section numerically verifies the validity of the four inequalities
presented in Sec. \ref{sec:2} using two examples.

\begin{figure}
\includegraphics[width=8cm,height=6cm]{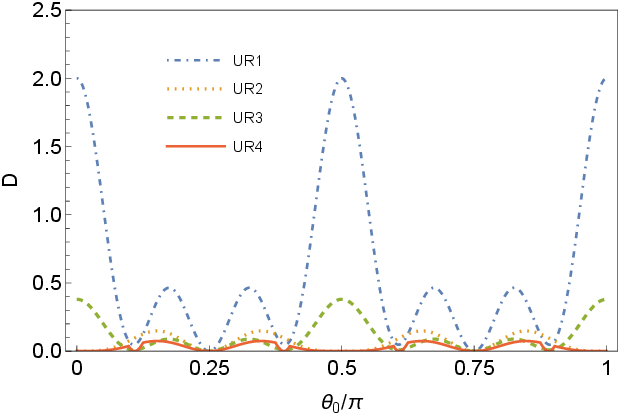}

\caption{\protect\label{fig:1}Difference $D$ between the LHS and RHS of four
distinct uncertainty relations as a function of $\theta_{0}$. The
dot-dashed curve denotes Eq. (\ref{eq:2}), the solid line curve is
for Eq. (\ref{eq:4}), the dashed curve for Eq. (\ref{eq:c}), and
the dotted curve for Eq. (\ref{eq:d}). It is considered that $\theta_{1}=\pi/4,\theta_{3}=\pi/3,\theta_{5}=\pi/4$,
and $\theta_{7}=3\pi/4$.}
\end{figure}

\textit{Example} 1: This example utilizes measurement in qubit systems.
In this first example, we only consider the non-Hermitian operators
and not the effect of $G$-metric formalism. Here, we follow the examples
illustrated of the experimental work of Ref. \citep{PhysRevLett.132.070203}.
We choose below two incompatible non-Hermitian operators: 

\begin{equation}
A=\left(\begin{array}{cc}
\frac{1}{2}\text{\ensuremath{\sin2\theta_{0}}} & \frac{1}{2}\cos2\theta_{0}\\
\cos2\theta_{0} & -\sin2\theta_{0}
\end{array}\right)\label{eq:38-1}
\end{equation}
\begin{equation}
B=\left(\begin{array}{cc}
0 & 0\\
\cos2\theta_{0}\  & -\sin2\theta_{0}
\end{array}\right),\label{eq:39-1}
\end{equation}
 and a qubit state parameterized by $\theta$ as 
\begin{equation}
\vert\psi\rangle=\cos2\theta_{0}\vert0\rangle+\sin2\theta_{0}\vert1\rangle,\label{eq:40-1}
\end{equation}
 where $\vert0\rangle$ and $\vert1\rangle$ are eigenstates of the
Pauli $z$ operator $\sigma_{z}$ corresponding to the eigenvalues
$1$ and $-1$, respectively, and $\theta_{0}\in[0,\pi]$. It is evident
that $A\neq A^{\dagger}$ and $B\neq B^{\dagger}$, and $\left[A,B\right]\neq0$
except for $\theta_{0}=\frac{\pi}{4}$ and $\frac{3\pi}{4}$. To check
the validity of Eqs. (\ref{eq:2}), (\ref{eq:4}), (\ref{eq:c}),
and (\ref{eq:d}) we need to compute their LHS and RHS separately.
Among them, the variances $\triangle A^{2}$and $\triangle B^{2}$
and other related average values can be calculated by using the above
information, and orthogonal states $\vert\psi_{A\pm B}^{\perp}\rangle$
and $\vert\psi_{A\pm B}^{\perp}\rangle$ also determined by the Aharonov--Vaidman
identity defined in Eq. (\ref{eq:A-V}). Our theoretical results are
validated on numerical analysis and shown in Fig. \ref{fig:1}. Figure
\ref{fig:1} illustrates the analytic numerical results, where the
horizontal axis is the angle $\theta$ required to prepare the initial
state $\vert\psi\rangle$ and the vertical axis is the difference
$D$ between the LHS and RHS of the URs corresponding to Eqs. (\ref{eq:2}),
(\ref{eq:4}), (\ref{eq:c}), and (\ref{eq:d}). Figure \ref{fig:1}
highlights that the four relations introduced are valid for all parameter
regions of state $\vert\psi\rangle$ except $\theta_{0}=\frac{\pi}{4},\frac{3\pi}{4}$.
Since at $\theta_{0}=\frac{\pi}{4}$,$\frac{3\pi}{4}$, the operators
$A$ and $B$ are commute, $[A,B]=0$, and then $D=0$ . This study
assumes that $A$ and $B$ are incompatible operators. Therefore,
the above two points are trivial cases. The closer the differences
$D$ approach 0, the tighter the bound. When comparing with other
bounds, the numerical results demonstrated that the particular bounds
of UR 3 and UR 4 outperform the competitor in specific regions of
the state space. Meanwhile, the bounds of UR1 and UR2 require optimization
over the states' parameters. Next, we briefly explain the preparation
of the above two non-Hermitian of operators $A$ and $B$ in the Laboratory.

The spin-$\frac{1}{2}$ qubit system is cornerstone of quantum information
sciences and its manipulation techniques are mature in the Laboratory.
As illustrated in Ref. \citep{PhysRevLett.132.070203} the realization
of above two real non-Hermitian operators can be accomplished by optical
components such as the use of half-wave plates (HWP) and quarter-wave
plates (QWP). Any non-Hermitian operator $A$ and $B$ can be expressed
as $A=S_{A}U_{A}$ and $B=S_{B}U_{B}$, respectively, using polar
decomposition \citep{PhysRevA.92.052120}, where $S=\sqrt{AA^{\dagger}}$
is a positive-semidefinite operator and $U$ is the corresponding
unitary operator. Following Ref. \citep{PhysRevLett.132.070203},
the polar decomposition parts of the non-Hermitian operators $A$
and $B$ are

\begin{align}
U_{A} & =\left[\begin{array}{cc}
\cos2\left(\theta_{1}-\theta\right) & \sin2\left(\theta_{1}-\theta\right)\\
\sin2\left(\theta_{1}-\theta\right) & -\cos2\left(\theta_{1}-\theta\right)
\end{array}\right],\label{eq:32}\\
S_{A} & =\left[\begin{array}{cc}
-\cos2\theta_{3} & 0\\
0 & 1
\end{array}\right],\label{eq:33}\\
U_{B} & =\left[\begin{array}{cc}
\cos2\left(\theta_{5}-\theta\right) & \sin2\left(\theta_{5}-\theta\right)\\
\sin2\left(\theta_{5}-\theta\right) & -\cos2\left(\theta_{5}-\theta\right)
\end{array}\right],\label{eq:34}\\
S_{B} & =\left[\begin{array}{cc}
-\cos2\theta_{7} & \ 0\\
0 & 1
\end{array}\right].\label{eq:35}
\end{align}

As investigated in Ref. \citep{PhysRevLett.132.070203}, these real
non-Hermitian operators (both $A$ and $B$) can be realized by a
phase-adjustable Sagnac ring interferometer and beam displacer (BD)
crystals. From the experimental point of view, $\theta_{k}$ can denoted
as the angle of the $k$-th half wave plate. The operators $A$ and
$B$ can be given with the above decomposition parts if we choose
$\theta_{1}=\pi/4$, $\theta_{3}=\pi/3$, $\theta_{5}=\pi/4$, and
$\theta_{7}=3\pi/4$ in the experimental work \citep{PhysRevLett.132.070203}.
In the optical experiment, the qubit basis vector $\vert0\rangle=[1,0]^{T}$
and $\vert1\rangle=[0,1]^{T}$ can be characterized by the horizontal
and the vertical polarization of a photon, i.e., $\vert H\rangle=\vert0\rangle$
and $\vert V\rangle=\vert1\rangle$. 

\textit{Example} 2: The first example demonstrates the validity of
the four improved URs proposed in Sec. \ref{sec:2} in a general non-Hermitian
system. The second example aims to confirm the validity of these four
sum URs within the $G$-inner product and good observables, considering
both the $\mathcal{PT}$ broken and unbroken phases. This example
is based on the simplest one-parameter $\mathscr{\mathcal{PT}}$-invariant
non-Hermitian system described by the Hamiltonian \citep{PhysRevLett.89.270401}

\begin{equation}
H\left(\gamma\right)=\sigma_{x}+i\gamma\sigma_{z}=\left(\begin{array}{cc}
i\gamma & 1\\
1 & -i\gamma
\end{array}\right),\label{eq:36}
\end{equation}
where $\gamma$ represents the non-Hermitian degree. The Hamiltonian
varies continuously with parameter $\gamma$, and adjustments to this
real parameter influence the $\mathcal{PT}$-symmetry of the system.
Specifically, when $\gamma^{2}<1$, the system resides within the
$\mathcal{PT}$ preserving region. Accordingly, for $\gamma^{2}>1$,
the system is in the $\mathcal{PT}$ breaking region. The eigenvalues
are $E_{1,2}=\pm\sqrt{1-\gamma^{2}}$ signifying the occurrence of
a phase transition precisely at the exceptional point (EP) $\gamma^{2}=1$.
The real parameter $\gamma$ also quantifies the strength of gain
and loss (diagonal) terms compared to the interlevel interactions.
Furthermore, the non-Hermitian Hamiltonian $H\left(\gamma\right)$
is a good observable in the $\mathcal{PT}$-symmetric phase but not
in the $\mathcal{PT}$-broken phase, suggesting the existence of an
EP in the system. The EP in a non-Hermitian system refers to a unique
point in the parameter space where both eigenvalues and eigenvectors
merge into a single value and state \citep{2012Heiss}. The two-level
non-Hermitian system characterized by Hamiltonian $H(\gamma)$ given
in Eq. (\ref{eq:36}) is widely investigated in various systems including
optics \citep{RN5,RN9}, ultracold atoms \citep{RN8}, and open quantum
systems \citep{2009Rotter,RN10}.

The proposed scheme assumes that the arbitrary initial state is prepared
into a general superposition of the eigenstates of the Hamiltonian
$H(\gamma)$ in the $\mathcal{PT}$-symmetric phases ($\gamma^{2}<1$),
expressed as

\begin{equation}
\vert\Psi\rangle=\mathscr{N}\left(\vert E_{1}\rangle+pe^{i\alpha}\vert E_{2}\rangle\right),\label{eq:37}
\end{equation}
where $p$ and $\alpha$ are real parameters and $\mathscr{N}$ is
the normalization coefficient. In this $\mathcal{PT}$-symmetric phase
region the subsystems are strongly coupled and the eigenvalues $E_{1.2}$
become real, and the entire system is in equilibrium. Additionally,
the eigenstates oscillate and do not grow or decay. The initial state
is set to be normalized for the $G$-inner product $\langle\langle\Psi\vert\Psi\rangle\rangle=\langle\Psi\vert G\vert\Psi\rangle=1$.
The right eigenstates of $H(\gamma)$ in the $\mathcal{PT}$preserving
region of the Hamiltonian, i.e., $H\left(\gamma\right)\vert E_{i}\rangle=E_{i}\vert E_{i}\rangle$
are

\begin{align}
\vert E_{1}\rangle= & \frac{1}{\sqrt{2\cos\theta}}\left[\begin{array}{c}
e^{i\theta/2}\\
e^{-i\theta/2}
\end{array}\right],\label{eq:38}\\
\vert E_{2}\rangle= & \frac{i}{\sqrt{2\cos\theta}}\left[\begin{array}{c}
e^{-i\theta/2}\\
-e^{i\theta/2}
\end{array}\right].\label{eq:39}
\end{align}
In this context, $\cos\theta=\sqrt{1-\gamma^{2}}$. The $\vert E_{1}\rangle$
and $\vert E_{2}\rangle$ are the right eigenvalues of $H$, and the
matrix of $G$ for this system can be determined by $\sum_{i}\vert E_{i}\rangle\langle E_{i}\vert G_{s}=1$
as $G_{s}=\left[\sum_{i}\vert E_{i}\rangle\langle E_{i}\vert\right]^{-1}$.
Thus, for the $\mathcal{PT}$-unbroken phase, the matrix of $G$ can
be chosen as follows at time $t=0$, despite $G$ not being uniquely
defined for a given $H(\gamma)$ \citep{PhysRevA.100.062118} 

\begin{equation}
G_{s}=\frac{1}{\sqrt{1-\gamma^{2}}}\left(\begin{array}{cc}
1 & -i\gamma\\
i\gamma & 1
\end{array}\right).\label{eq:40}
\end{equation}
Similarly, in the $\mathcal{PT}$-broken phases, the normalized right
eigenstates of $H(\gamma)$ are

\begin{align}
\vert e_{1}\rangle= & \frac{1}{\sqrt{2\gamma\lambda-2\lambda^{2}}}\left[\begin{array}{c}
1\\
-i\left(\gamma-\lambda\right)
\end{array}\right],\label{eq:41}\\
\vert e_{2}\rangle= & \frac{1}{\sqrt{2\gamma\lambda-2\lambda^{2}}}\left[\begin{array}{c}
i\left(\gamma-\lambda\right)\\
1
\end{array}\right],\label{eq:42}
\end{align}
where $\lambda=\sqrt{\gamma^{2}-1}$. In this case, $\gamma^{2}>1$.
In this $\mathcal{PT}$-broken phase region the subsystems are weakly
coupled, the eigenvalues $E_{1.2}$ are complex , and the system is
not in equilibrium, i.e., one eigenstate grows in time and the other
decays in time.

\begin{figure}
\includegraphics[width=8cm,height=6cm]{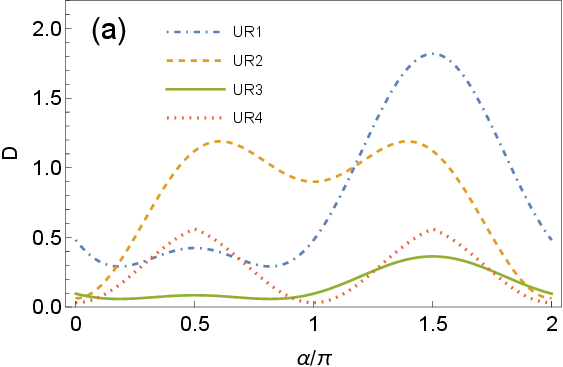}\\
\vspace{2cc}

\includegraphics[width=8cm,height=6cm]{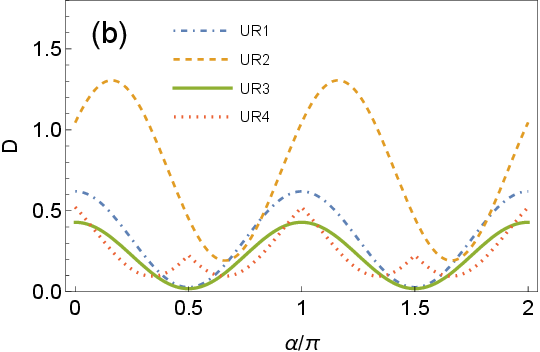}

\caption{\protect\label{fig:2}Difference $D$ between the LHS and RHS for
four distinct URs as a function of state parameter $\alpha$. The
dot-dashed curves refer to Eq. (\ref{eq:G1b}), the solid line curves
to Eq. (\ref{eq:14}), the dashed curves to Eq. (\ref{eq:30}), and
the dotted curves to Eq. (\ref{eq:31}), which are plotted for two
non-Hermitian systems in the second example, in the range of $0\protect\leq\alpha\protect\leq2\pi$.
(a) Inequalities for two good observables $H\left(\gamma\right)$
and $\sigma_{y}$ in $\mathcal{PT}$-symmetric phase ($\gamma=0.9$,
$p=0.5$). (b) Inequalities for two good observables $H\left(1/\gamma\right)$
and $\sigma_{y}$ in $\mathcal{PT}$-broken phase ($\gamma=1.2$,
$p=1.5$). }
\end{figure}

We assume that for this $\mathcal{PT}$-broken phase region, the corresponding
arbitrary initial state is $\vert\Phi\rangle=\mathscr{N}\left(\vert e_{1}\rangle+pe^{i\alpha}\vert e_{2}\rangle\right)$.
By using $G_{b}=\left[\sum_{i}\vert e_{i}\rangle\langle e_{i}\vert\right]^{-1}$,
the matrix of $G_{b}$ for the $\mathcal{PT}$ broken phase is:

\begin{align}
G_{b} & =\frac{1}{\sqrt{\gamma^{2}-1}}\left(\begin{array}{cc}
\gamma & -i\\
i & \gamma
\end{array}\right).\label{eq:43}
\end{align}
Notably, selecting good observables in non-Hermitian systems depends
on the $G$ metric, exhibiting different characteristics in $\mathcal{PT}$-symmetric
and $\mathcal{PT}$-broken phases. In this example, $H\left(\gamma\right)$
is effective as a good observable in the unbroken phase but not in
the broken phase. Hence, to illustrate the effectiveness of the proposed
URs, we choose two incompatible good observables, $H\left(\gamma\right)$
and $\sigma_{y}$ for the $\mathcal{PT}$ unbroken phase, with the
numerical results depicted in Fig. \ref{fig:2} (a). Considering the
$\mathcal{PT}$-broken phase, we choose $H\left(1/\gamma\right)$
and $\sigma_{y}$ as good observables. Figure \ref{fig:2} (b) depicts
the differences $D$ of LHS and RHS of our sum URs for that region.
As indicated in Fig. \ref{fig:2}, our sum URs hold well for all parameter
regions.

In both regions, to illustrate the preceding uncertainty relations,
it is necessary to identify the state $\vert\psi^{\perp}\rangle$,
which is orthogonal to the system state $\vert\psi\rangle$. In the
non-Hermitian system, the orthogonality is expressed through the $G$-inner
product $\langle\psi^{\perp}\vert G\vert\psi\rangle=\langle\langle\psi^{\perp}\vert\psi\rangle=0$.
Employing eigenvectors that constitute a complete biorthogonal set
between left and right eigenvector $\langle L_{i}\vert$ and $\vert R_{j}\rangle$
which satisfy $\langle L_{i}\vert R_{j}\rangle=\delta_{ij}$, and
allow designating the corresponding left eigenvectors $\langle L_{i}\vert$
of the Hamiltonian to the state $\langle\langle\psi^{\perp}\vert$.
Moreover, based on the definition derived from the Aharonov-Vaidman
identity, $\vert\psi_{A\pm B}^{\perp}\rangle$ should be modified
as

\begin{equation}
\vert\psi_{A\pm B}^{\perp}\rangle_{G}=\frac{(A\pm B)-\langle A\pm B\rangle_{G}}{\Delta(A\pm B)_{G}}\vert\psi\rangle.\label{eq:44}
\end{equation}
In our analysis, the difference between the LHS and RHS of the URs
is defined as $D$. When $D=0$, the uncertainty relation is satisfied
with equality, corresponding to a minimum UR. The $D$ for two incompatible
observables is plotted to compare the four proposed URs under the
modified NHQM while using separately the Hermitian $G$ metric operator
for $\mathcal{PT}$-symmetric and broken phases. For the second example,
a comparison is made in Fig. \ref{fig:2}, with the plots revealing
that the lower bound of the third and fourth URs are tighter than
the others for the $\mathcal{PT}$-unbroken and $\mathcal{PT}$-broken
religions. When $\alpha=\pm1$, the first UR given by Eq. (\ref{eq:G1b})
implies that one has $\triangle A_{G}^{2}+\triangle B_{G}^{2}\geq\pm i\langle\psi\vert G\left[A,B\right]\vert\psi\rangle$,
while the third UR given by Eq. (\ref{eq:30}) is stronger, and the
numerical examples can test this. Furthermore, the plot reveals that
the bound of UR3 is consistently tighter than UR1, even when adjusting
other parameters.

\section{\protect\label{sec:4} Discussion and outlook}

This section focuses on the significance of correctly applying the
NHQM formalism for the URs on non-Hermitian $\mathcal{PT}$-symmetric
systems.

Assume that $A$ and $B$ are Hermitian operators, e.g., $A^{\dagger}=A$
and $B^{\dagger}=B$, the corresponding forms of our first and second
sum URs in CQM can be written as 

\begin{equation}
\Delta A^{2}+\Delta B^{2}\geq i\langle\left[B,A\right]\rangle,\label{eq:CQM=0000201}
\end{equation}
 and 

\begin{equation}
\Delta A^{2}+\Delta B^{2}\geq\langle\{A,B\}\rangle-2\langle A\rangle\langle B\rangle.\label{eq:CQM=0000202}
\end{equation}
The above two expressions can be easily obtained using the Hermitian
condition of operators in our derivations introduced in Sec. \ref{sec:2}.
Referring to Table \ref{table:1}, it becomes evident that through
correctly applying the NHQM formalism with $G$-metric formalism,
the derived uncertainty relations exhibit a similar structure to Eqs.
(\ref{eq:CQM=0000201}) and (\ref{eq:CQM=0000202}) in CQM.

Furthermore, in the CQM realm, the correspondent of the third and
fourth URs are expressed as \citep{PhysRevLett.113.260401} 

\begin{align}
\triangle A^{2}+\triangle B^{2} & \geq\pm i\langle\psi\vert\left[A,B\right]\vert\psi\rangle+\left|\langle\psi\vert A\pm iB\vert\psi^{\perp}\rangle\right|^{2},\label{eq:49}\\
\triangle A^{2}+\triangle B^{2} & \geq\frac{1}{2}\left|\langle\psi_{A+B}^{\perp}\vert A+B\vert\psi\rangle\right|^{2}.\label{eq:50}
\end{align}
Comparing the above two relations with the findings presented in Sec.
\ref{sec:2} for the NHQM realm {[}refer to Eqs. (\ref{eq:c}) and
(\ref{eq:d-1}){]}, it is found that they are different from the conventional
ones. However, in the $G$-metric formalism, the above-modified relations
for non-Hermitian operators within a special inner product framework
{[}refer to Eqs. (\ref{eq:30}) and (\ref{eq:31}){]} have a similar
form to the conventional Hermitian expressions.

It is evident that if $G=1$ (Dirac inner product), the UR 1 to 4
trivially reduced to Hermitian-type correspondences. As discussed
above, the generalized expression of uncertainty relations in a non-Hermitian
system differs from the conventional ones. Hence, it is necessary
to provide the corresponding modifications for the relations using
a proper Hilbert-space metric.

We discuss the nontrivial lower bounds of the four sum URs in NHQM.
It is clear that $\Delta X^{2}$ depends on the state of the system
$\vert\psi\rangle,$ for a non-Hermitian operator $X$, and $\Delta X^{2}=0$,
if $\vert\psi\rangle$ is an eigenvector of $X$ \citep{2013Brian}.
An interesting case is if state $\vert\psi\rangle$ is an eigenvector
of either $A$ or $B$, and the lower bound of the first and second
URs can be zero. For the third UR, the lower bound depends on the
incompatibility of the two non-Hermitian operators on the special
states and the optimization of $\vert\psi^{\perp}\rangle$. By optimizing
$\vert\psi^{\perp}\rangle$ and the corresponding initial state, the
third UR can be saturated to equality. For example, if $\vert\psi\rangle$
is an eigenstate of $B$, then $\vert\psi^{\perp}\rangle=\frac{\left(A-\langle A\rangle\right)\vert\psi\rangle}{\Delta A}$
should be chosen to maximize the lower bound, where both sides of
the inequality become $\Delta A^{2}$. Moreover, the lower bound can
be nonzero even if the state $\vert\psi\rangle$ is an eigenvector
of $A$ ($B$), i.e., just choose $\vert\psi^{\perp}\rangle$ orthogonal
to $\vert\psi\rangle$ but not orthogonal to the state $A\vert\psi\rangle$
($B\vert\psi\rangle$) . If the incompatible operators lack a common
eigenstate, the fourth UR will have a nontrivial bound, except for
the trivial case when $\vert\psi\rangle$ is an eigenstate of $A\pm B$.
The form of $\vert\psi_{A\pm B}^{\perp}\rangle$ implies that even
if $\vert\psi\rangle$ is an eigenvector of $A$ or $B$, the RHS
always yields a nonzero value as $\frac{1}{2}\triangle A^{2}$ or
$\frac{1}{2}\triangle B^{2}$, respectively.

Next, we delve into the nontrivial lower bounds of four modified sum
URs within the $G$-metric formalism. If $\vert\psi\rangle$ is an
eigenvector of $A$ or $B$, constraining the condition of $\mathcal{PT}$-symmetry
on the other operator, the RHS of first and second URs are zero. Otherwise,
the RHS will be proportional to the imaginary part of the eigenvalue
of the corresponding operator for the first and second URs. For the
same case, if it is considered to optimize the orthogonal states,
the third UR can be transformed into an equality. For example, if
$\vert\psi^{\perp}\rangle$ is an eigenstate of $B$, by choosing
$\vert\psi^{\perp}\rangle=\frac{\left(A-\langle GA\rangle\right)\vert\psi\rangle}{\Delta A_{G}}$,
the RHS becomes $\pm2Im\left(E_{B}^{*}\right)\langle GA\rangle+\Delta A_{G}^{2}$,
which maximizes the lower bound for the $\mathcal{PT}$-symmetric
phase, where $E_{B}$ denotes the eigenvalue of $B$ operator in state
$\vert\psi\rangle$. While in the $\mathcal{PT}$-broken phase, unless
there is a common eigenstate of $A$ and $B$, the above term reveals
a remnant uncertainty in the lower bound. For the fourth UR, in the
example where $\vert\psi\rangle$ is the eigenvalue of $B$, the RHS
is $\max\left[\frac{1}{2}\left|\frac{\langle AGA\rangle_{G}-\langle GA\rangle_{G}\langle A\rangle_{G}\pm E_{B}\left(\langle AG\rangle_{G}-\langle G\rangle_{G}\langle A\rangle_{G}\right)}{\triangle A_{G}}\right|^{2}\right]$.
Suppose the operators do not share a common eigenstate. In this case,
the lower bound shows a nonzero amount of uncertainties, even if the
state is the eigenstate of one of the operators in both the $\mathcal{PT}$-symmetric
and $\mathcal{PT}$-broken phases.

Regarding the dimensionality of operators in the sum uncertainty relation,
we need to handle it with caution when the two operators involved
have different dimensions. From a mathematical perspective, if the
operators have different dimensions, their sum may not have a well-defined
physical meaning, unless they are additive in a specific physical
context. To address this issue, one approach is to construct dimensionless
combinations of the operators involved. Another method introduces
the appropriate scaling factors to help reconcile the dimensional
differences. For instance, if one operator has units of length and
the other has units of momentum, we could define a new operator that
incorporates a scaling factor, effectively making their dimensions
compatible. However, as\textcolor{blue}{{} }investigated in previous
works \citep{PhysRevLett.113.260401,PhysRevA.93.052108}, compared
with the position and momentum operators, which have different physical
dimensions, the spin-$\frac{1}{2}$ \citep{2015Chen,PhysRevA.96.062123,2019Chen}
and spin-1 \citep{PhysRevA.98.032118} physical systems with $N$
non-commuting operators ($N\ge2$) are ideal platforms to test the
sum URs. Since non-commuting operators have the same physical dimension
in the spin physical systems, the challenging re-scaling processes
of different dimensional operators can be avoided. Furthermore, as
introduced in Ref. \citep{PhysRevA.68.032103}, it is also possible
to apply the sum URs where the two Hilbert spaces of systems $A$
and $B$ do not need to have the same dimension.

\begin{widetext}

\begin{table}
\caption{\protect\label{table:1} Four sum URs in NHQM (left) and their modified
forms with $G$-metric formalism (right).}

\begin{tabular}{|c|c|c|}
\hline 
 & Non-Hermitian quantum mechanics & Modified Non-Hermitian quantum mechanics\tabularnewline
\hline 
\hline 
UR 1 & $\Delta A^{2}+\Delta B^{2}\geq2Im\left[Cov\left(A,B\right)\right]$ & $\Delta A_{G}^{2}+\Delta B_{G}^{2}\geq i\langle\left[B,A\right]\rangle_{G}$\tabularnewline
\hline 
UR 2 & $\Delta A^{2}+\Delta B^{2}\geq2Re\left[Cov\left(A,B\right)\right]$ & $\Delta A_{G}^{2}+\Delta B_{G}^{2}\geq\langle\left\{ A,B\right\} \rangle_{G}-2\langle B\rangle_{G}\langle A\rangle_{G}$\tabularnewline
\hline 
UR 3 & $\triangle A^{2}+\triangle B^{2}\geq\pm2Im\left[Cov\left(A,B\right)\right]+\left|\langle\psi^{\perp}\vert\left(A\pm iB\right)\vert\psi\rangle\right|^{2}$ & $\triangle A_{G}^{2}+\triangle B_{G}^{2}\geq\pm i\langle\left[A,B\right]\rangle_{G}+\left|\langle\psi\vert G\left(A\pm iB\right)\vert\psi^{\perp}\rangle\right|^{2}$\tabularnewline
\hline 
UR 4 & $\Delta A^{2}+\Delta B^{2}\geq\max\left[\frac{1}{2}\left|\langle\psi_{A\pm B}^{\perp}\vert\left(A\pm B\right)\vert\psi\rangle\right|^{2}\right]$ & $\triangle A_{G}^{2}+\triangle B_{G}^{2}\geq\max\left[\frac{1}{2}\left|\langle\psi_{A\pm B}^{\perp}\vert G\left(A\pm B\right)\vert\psi\rangle\right|^{2}\right]$\tabularnewline
\hline 
\end{tabular}

\end{table}
\end{widetext}

In summary, the previous URs, which are based on the product of variances
of $\Delta A^{2}\Delta B^{2}$, do not fully capture the incompatibility
of the observables on the system state. In addition, the sum variances
of $\Delta A^{2}+\Delta B^{2}$ have not been given and modified in
non-Hermitian quantum mechanical systems within the $G$-metric formalism.
Directly applying the theorem and axioms of conventional quantum mechanics
to NHQM might cause some violations \citep{2014Pati,PhysRevA.90.054301,PhysRevLett.112.130404,2016Brody,2016Znojil}.
Therefore, this study establishes nontrivial lower bounds for the
sum of variances of two arbitrary incompatible operators, which apply
to the general non-Hermitian observables in NHQM. This study derives
four lower bounds of Eqs. (\ref{eq:2}), (\ref{eq:4}), (\ref{eq:c}),
and (\ref{eq:d}) for the sum variances, which are applicable when
the observables are incompatible concerning the system's state. In
addition, this research presents the tight bounds of those URs for
two good observables within the $G$-metric formalism in NHQM. Two
illustrative examples demonstrate the validity of the proposed sum
URs, highlighting that our four sum URs are adequate for all the parameter
regions of the given system states. We believe This paper can help
the reader understand the sum URs for NHQM in depth. Furthermore,
this study proves the usefulness of the $G$-metric in providing the
correct counterparts of formulas, theorems, and axioms of CQM into
the NHQM realm. 

Our sum URs have potential applications in quantum information theory,
such as entanglement detection \citep{PhysRevA.68.032103,PhysRevLett.92.117903},
testing error relation for joint measurements \citet{OZAWA2004367,PhysRevA.89.012129},
and measurement-induced disturbance \citep{PhysRevA.68.032103,PhysRevA.67.042105,RevModPhys.86.1261,PhysRevA.89.012129,2010Berta}
problems of non-Hermitian systems. In addition, our sum URs also may
be useful for security analysis of quantum cryptography \citep{PhysRevA.53.2038,PhysRevLett.103.020402}
based on non-Hermitian quantum systems. Although non-Hermitian physics
has achieved remarkable successes in practical applications, such
as in quantum open systems \citep{2009Rotter,2015Zhen,PhysRevLett.123.170401,2024Chen},
topology \citep{PhysRevLett.115.200402,PhysRevLett.121.086803,PhysRevB.99.235112,PhysRevX.9.041015},
non-reciprocal and chiral transport \citep{2021Fruchart}, metamaterials
\citep{PhysRevA.87.053824,PhysRevLett.112.143903,PhysRevLett.113.023903,PhysRevA.94.033834,2016Xiao},
and nonreciprocal device \citep{2011Liang,2014Peng}, this research
field still faces challenges in fully developing the theoretical framework
corresponding to CQM and its associated models. Significant theoretical
work remains to be done, and the introduction of the $G$-metric formalism
has established a bridge between NHQM and CQM. Nonetheless, experimental
verification of these theoretical developments of NHQM described by
considering the $G$-metric formalism may still require considerable
time and effort. Additionally, in experimental work \citep{2017Lei},
the sum UR for two unitary operators is successfully tested, and the
most recent experimental work by Guo $et$ $al$. \citep{PhysRevLett.132.070203}
suggests that sum URs for any kind of non-Hermitian operators could
be experimentally verified. Those experimental works also imply the
feasibility of measuring the expectation values of non-Hermitian Hamiltonians
including the $G$ metric. Therefore, it is anticipated that as the
theoretical framework of NHQM continues to evolve and associated experimental
techniques improve, the sum URs proposed in our current work also
will also be experimentally validated in the near future.
\begin{acknowledgments}
This work was supported by the National Natural Science Foundation
of China (No. 12365005)
\end{acknowledgments}

\bibliographystyle{apsrev4-1}
\bibliography{QCC-ref}

\end{document}